\documentstyle[aps]{revtex}
\begin{document}
\title{SPACE TIME DEFECTS AS A SOURCE OF CURVATURE AND TORSION}
\author{ANGELO TARTAGLIA}
\address{Dipartimento di Fisica del Politecnico and INFN, \\
Corso Duca degli Abruzzi 24, i-10129 Torino, Italy\\
e-mail: angelo.tartaglia@polito.it}
\maketitle

\begin{abstract}
Space time is described as a continuum four-dimensional medium similar to
ordinary elastic continua. Exploiting the analogy internal stress states are
considered. The internal "stress" is originated by the presence of defects.
The defects are described according to the typical Volterra process. The
case of a point defect in an otherwise isotropic four-dimensional medium is
discussed showing that the resulting metric tensor corresponds to an
expanding (or contracting) universe filled up with a non-zero
energy-momentum density.
\end{abstract}

\markboth{A. Tartaglia}{Curvature And Torsion From Space Time
Defects}

\section{Introduction}

The need of accounting for many data coming from the observation of the
universe at large scale is pushing many a scientist to introduce a number of
suspect entities, such ad dark matter, dark energy, ad hoc universal fields.
The scenario is rather unsatisfactory, resembling a bit to the one at the
end of the XIX century with its ether. Summing up all this it may be the
case to explore other schemes or paradigms, in quest at least of a
simplified and internally consistent description of what we already know.

The idea outlined in this paper is that the "physicity" of space-time may be
described as belonging to a sort of a medium that can be deformed and reacts
to strain more or less as ordinary materials do. An energy content in
space-time would be the manifestation of stress-strain induced by the
presence of defects, and finally all this would show up in the metric
properties of the same space-time. As we shall see, at the cosmological
scale the model appears to be reasonably consistent with the universe we see.

\section{Strain in the Continuum}

Let us start from an unstrained continuum in any number of dimensions. We
can label each point in the continuum with appropriate coordinates $x^{\mu }$%
, and, considering the analogy with ordinary three-dimensional homogeneous
unstrained media, we expect the geometry to be Euclidean. A strain
corresponds to a displacement field\cite{kroner} that moves each material
point to a new position, whose coordinates are obtained from the old ones
(in the old reference frame) as 
\begin{equation}
y^{\mu }=x^{\mu }+\xi ^{\mu }.  \label{displac}
\end{equation}
$\xi $'s are the components of the displacement vector field and in general
depend on $x$'s. If each displaced point drags with itself the coordinate
label, we may call the $x$'s "intrinsic" coordinates (the ones used by an
internal observer), while $y$'s taken with respect to the initial unstrained
background will be "extrinsic" coordinates.

Equation (\ref{displac}) is a diffeomorphism. In three-dimentional
continuous media, diffeomorphisms do not modify the geometry and in
particular do not introduce any intrinsic curvature. The strained state can
be maintained if the medium has a boundary and appropriate external forces
are applied at the boundary.

Bearing this important remark in mind we can go on determining what is the
metric tensor induced by the deformation field. Suppose the coordinates we
use are Cartesian ones. The unperturbed (Euclidean) metric tensor elements
will be $\delta _{\mu \nu }$ (Kronecker deltas). The induced metric is then: 
\begin{equation}
g_{\mu \nu }=\delta _{\alpha \beta }\frac{\partial y^{\alpha }}{\partial
x^{\mu }}\frac{\partial y^{\beta }}{\partial x^{\nu }}.  \label{induced}
\end{equation}

Considering Eq. (\ref{displac}) the induced metric can be written 
\begin{equation}
g_{\mu \nu }=\delta _{\mu \nu }+2\varepsilon _{\mu \nu },  \label{metrica}
\end{equation}
where the strain tensor $\varepsilon _{\mu \nu }$ has been introduced, whose
explicit expression is 
\begin{equation}
\varepsilon _{\mu \nu }=\frac{1}{2}\left( \frac{\partial \xi ^{\mu }}{%
\partial x^{\nu }}+\frac{\partial \xi ^{\nu }}{\partial x^{\mu }}+\delta
_{\alpha \beta }\frac{\partial \xi ^{\alpha }}{\partial x^{\nu }}\frac{%
\partial \xi ^{\beta }}{\partial x^{\mu }}\right).  \label{strain}
\end{equation}
In the linear theory of elasticity the non-linear terms in the brackets are
neglected.

The general De Saint-Venant integrability conditions\cite{dsv} for the six
differential equations (\ref{induced}), $R_{jkl}^{i}=0$, reduce, in three
dimensions, to the six independent conditions 
\begin{equation}
R_{ij}=0,  \label{ricci}
\end{equation}
where $R$'s are the components of the Ricci tensor.

\section{Defects}

The internal state of a strained medium assumes a non-trivial aspect when
introducing the notion of a "defect". The way I shall describe defects is
the same, effective and vivid, introduced by Volterra\cite{volterra} at the
beginning of the XX century, studying the properties of crystals. Suppose to
cut away an remove a piece of your material unstrained continuum, then close
the hollow gluing together the corresponding surfaces of the cut. The final
state will be a defect; in the bulk we shall find stresses which stay there
without need for peculiar boundary conditions: they are induced by the
presence of the defect.

The theory of defects is a well established one in three dimensions\cite
{kron} and a complete classification exists of the different possibilities;
I am not going to review this theory here, but I would like to extend its
reach to four dimensions (in principle even more), showing that it can be
more than a formal exercise.

\section{A Point Defect in Four Dimensions}

Let me consider a vacancy, or, otherwise frased, a point defect. This is the
topologically simplest situation for a defect. Applying the Volterra process
we may think of removing a whole sphere from the isotropic and homogeneous
medium, then close the hole pushing its surface radially in.

Radial stresses and strains will result in the bulk. To see what happens in
four dimensions let us begin with an Euclidean unstrained medium described
by means of a four-dimensional polar coordinate system. The line element is
of course: 
\begin{equation}
ds^{2}=d\rho ^{2}+\rho ^{2}\left( d\psi ^{2}+\sin ^{2}\psi \left( d\theta
^{2}+\sin ^{2}\theta d\phi ^{2}\right) \right).  \label{four-line}
\end{equation}

The removal of a sphere etc., as described above, introduces a radial
displacement field. Making use of the analogy with the corresponding
three-dimensional elastic problem and as far as the linear theory of
elasticity works, i.e. Hooke's law is applicable, the solution for the
displacement field is: 
\begin{equation}
\xi ^{\rho }=\frac{Q}{\rho },  \label{radiale}
\end{equation}
where $Q$ is a constant. It is $Q<0$ when the movement is inwards, $Q>0$ in
the opposite case.

Of course (\ref{radiale}) diverges in the origin, which cannot be the case.
The maximum displacement is the one for points on the surface of the sphere,
and equals, of course, its radius $R$. The simplest guess for a functional
form respecting the constraint of being finite in the origin and going
smoothly to (\ref{radiale}) is 
\begin{equation}
\xi ^{\rho }=\frac{Q}{k+\rho },  \label{displace}
\end{equation}
where $k$ is a positive constant and 
\begin{equation}
R=\left\vert \frac{Q}{k}\right\vert.  \label{raggio}
\end{equation}

The displacement vector field is singular in the origin although the modulus
of the vector is everywhere finite. According to the elastic analogy the
stresses also will diverge in the origin, and any linear approximation (then
Hooke's law) will cease to be acceptable while approaching the vacancy.

In a sense $R$ represents the "strength" of the defect, whose presence is
testified by the non vanishing integral 
\begin{equation}
\int_{0}^{\infty }\frac{\partial \xi ^{\rho }}{\partial \rho }d\rho =-R.
\label{integrale}
\end{equation}

\subsection{The induced metric tensor}

From (\ref{metrica}), (\ref{displace}), and (\ref{strain}) we deduce the
induced metric tensor. The only affected element is 
\begin{equation}
g_{\rho \rho }=\left( 1+\frac{\partial \xi ^{\rho }}{\partial \rho }\right)
^{2}=\left( 1-\frac{Q}{\left( k+\rho \right) ^{2}}\right) ^{2}  \label{groro}
\end{equation}
and the whole line element is 
\begin{equation}
ds^{2}=\left( 1-\frac{Q}{\left( k+\rho \right) ^{2}}\right) ^{2}d\rho
^{2}+\rho ^{2}d\psi ^{2}+\rho ^{2}\sin ^{2}\psi \left( d\theta ^{2}+\sin
^{2}\theta d\phi ^{2}\right).  \label{lineaeu}
\end{equation}

\section{A Vacancy in Space-Time}

Our unperturbed initial state was Euclidean in order to have a more direct
correspondence with the behaviour of three-dimensional elastic media, but it
is now time to extend the method to actual space-time. The "trick" to
convert the results obtained from the Euclidean four-dimensional case into
analogous ones based on an unperturbed Minkowski space-time is to transform
the ordinary angle $\psi $ into the hyperbolic angle $\chi =i\psi $.

The resulting new line element will be 
\begin{equation}
ds^{2}=\left( 1-\frac{Q}{\left( k+\rho \right) ^{2}}\right) ^{2}d\rho
^{2}-\rho ^{2}d\chi ^{2}-\rho ^{2}\sinh ^{2}\chi \left( d\theta ^{2}+\sin
^{2}\theta d\phi ^{2}\right).  \label{mink}
\end{equation}

Let us perform the simple coordinate change: 
\begin{equation}
\rho +\frac{Q}{k+\rho }=\tau +\frac{Q}{k}.  \label{cochange}
\end{equation}
The line element becomes 
\begin{eqnarray}
ds^{2} &=&d\tau ^{2}-a^{2}\left( d\chi ^{2}+\sinh ^{2}\chi \left( d\theta
^{2}+\sin ^{2}\theta d\phi ^{2}\right) \right),  \nonumber \\
a &=&\frac{1}{2k}\left( Q-k^{2}+k\tau +\sqrt{2Qk\tau +\left( k^{2}-Q\right)
^{2}+2k^{3}\tau +k^{2}\tau ^{2}}\right).  \label{univ}
\end{eqnarray}

Equation (\ref{univ}) coincides with a cosmic hyperbolic solution of the
Einstein's equations. The scale factor is $a$. Space is conformally flat,
and the expansion is an accelerated one ($\stackrel{\cdot \cdot }{a}\neq 0$).

\subsection{Stress-energy content}

A question at this point is: what is the source for the metric (\ref{univ}).
The answer is, in a sense, standard. Calculating the Einstein tensor $G_{\nu
}^{\mu }$ we obtain an energy density 
\begin{equation}
w=-\frac{c^{2}}{16\pi G}G_{0}^{0}=-\frac{3c^{2}}{16\pi G}\frac{\stackrel{%
\cdot }{a}^{2}-1}{a^{2}}  \label{energia}
\end{equation}
and a pressure 
\begin{equation}
\frac{16\pi G}{c^{2}}p=-G_{1}^{1}=-G_{2}^{2}=-G_{3}^{3}=\frac{1-2a\stackrel{%
\cdot \cdot }{a}+\stackrel{\cdot }{a}^{2}}{a^{2}}.  \label{pressione}
\end{equation}

Our defected space-time behaves like a perfect fluid, whose stress-energy
content is due to the strain caused by the "initial" vacancy.

The space-time curvature is 
\[
\kappa =6\frac{1-a\stackrel{\cdot \cdot }{a}+\stackrel{\cdot }{a}^{2}}{a^{2}}%
. 
\]

\section{Preliminary Conclusions}

We have seen that a simple analogy between space-time and a physical
continuum containing defects provides a framework able to reproduce or at
least re-interpret known results regarding the evolution of the universe at
large. Ours of course is a physical empty space-time. One should now
introduce matter. Pushing the analogy further, we could think that matter
appears in the form of other linear defects, meaning by "linear" that a
classical massive object corresponds to a time-like world-line. The
interaction among masses would be mediated by the additional strains
introduced by the presence of these new "defects".

Considering more elaborated types of defects it is also possible to directly
introduce consequences at the level of the connection and in particular to
introduce torsion.\cite{punti}

Of course what we have drawn in the few lines above is rather a programme
than a conclusion. The idea seems promising and, in the least, we have
introduced a simple paradigm to account for dark energies, quintessence
fields, and other similar entities.


\begin{references}
\bibitem{kroner}  H. Kleinert, in {\it Gauge Fields in Condensed Matter}, 
{\it Vol. II:} {\it Stresses and Defects }(World Scientific, Singapore,
1989).

\bibitem{dsv}  A. J. C. B. St. Venant, in {\it R\'{e}sum\'{e} des Le\c{c}ons
sur l'application de la M\'{e}canique}, ed. C.L.M.H. Navier (Didot, Paris,
1864).

\bibitem{volterra}  V. Volterra, {\it Ann. \'{E}c. Norm. Sup.} {\bf 24}, 401
(1907).

\bibitem{kron}  E. Kr\"{o}ner, in {\it Les Houches 1980, Session XXXV:
Physics of Defects}, ed. R. Balian, M. Kl\'{e}man, and J. P. Poirier, (North
Holland, Amsterdam, 1980), pp. 215--315.

\bibitem{punti}  R. A. Puntigam and H. H. Soleng, {\it Class. Quantum Grav.} 
{\bf 14}, 1129 (1997).
\end{references}
\end{document}